\documentclass{article}

\usepackage{arxiv}

\usepackage[utf8]{inputenc} 
\usepackage[T1]{fontenc}    
\usepackage{hyperref}       
\usepackage{url}            
\usepackage{booktabs}       
\usepackage{amsfonts}       
\usepackage{nicefrac}       
\usepackage{microtype}      
\usepackage{lipsum}		
\usepackage{graphicx}
\usepackage{natbib}
\usepackage{doi}

\usepackage{amssymb}
\usepackage{amsmath}
\usepackage{amsthm}
\usepackage{tabularx}
\usepackage{multirow}
\usepackage{array}
\usepackage{cases}
\usepackage{txfonts}
\usepackage{mathrsfs}
\usepackage{graphicx}
\usepackage{adjustbox}
\usepackage{subfigure}
\usepackage[]{hyperref}
\usepackage{geometry}
\usepackage{setspace}
\usepackage{enumitem}
\usepackage{color}
\usepackage{soul}

\usepackage[figuresright]{rotating}
\usepackage{makecell}

\title{A Survey on Deep Neural Network Partition over Cloud, Edge and End Devices}

\author{\hspace{1mm}Di Xu, Xiang He, Tonghua Su, Zhongjie Wang \\
	Faculty of Computing\\
	Harbin Institute of Technology\\
	Harbin, China 150001 \\
	\texttt{xudi@stu.hit.edu.cn, \{hexiang, thsu, rainy\}@hit.edu.cn}
}

\renewcommand{\shorttitle}{A Survey on Deep Neural Network Partition over Cloud, Edge and End Devices}

\begin{document}
\maketitle

\begin{abstract}
``Deep neural network (DNN) partition'' is a research problem that involves splitting a DNN into multiple parts and offloading them to specific locations. Because of the recent advancement in  multi-access edge computing and edge intelligence, DNN partition has been considered as a powerful tool for improving DNN inference performance when the computing resources of edge and end devices are limited and the remote transmission of data from these  devices to clouds is costly. This paper provides a comprehensive survey on the recent advances and challenges in DNN partition approaches over the cloud, edge, and end devices based on a detailed literature collection. We review how DNN partition works in various application scenarios, and provide a unified mathematical model of the DNN partition problem. We developed a five-dimensional classification framework for DNN partition approaches, consisting of deployment locations, partition granularity, partition constraints, optimization objectives, and optimization algorithms. Each existing DNN partition approache can be perfectly defined 
in this framework by instantiating each dimension into specific values. In addition, we suggest a set of metrics for comparing and evaluating the DNN partition approaches. Based on this, we identify and discuss research challenges that have not yet been investigated or fully addressed. We hope that this work helps DNN partition researchers by highlighting significant future research directions in this domain.
\end{abstract}


\section{Introduction}
\label{}
\subsection{Background} 
Deep neural networks (DNNs) have achieved considerable success in various machine-learning applications in recent years. The DNN model is a specific type of artificial neural network with multiple layers of feature extraction~\citep{ashouri2020analyzing}. It has consistently achieved state-of-the-art performance on various tasks, such as computer vision, natural language processing, intelligent personal assistance services, augmented reality, smart homes, and smart cities~\citep{gu2021survey,schmidhuber2015deep}. 
Because of the rapid spread of Internet of Things (IoT) devices (e.g., wearable sensors) that are integrated into all aspects of people’s lives, researchers are aiming to study more complex DNNs with high accuracy~\citep{he2015delving}. However,as DNNs become deeper or more complex~\citep{szegedy2015going}, they require higher processing capabilities to achieve an acceptable latency for training and inference, including the requirements of energy, memory, processors, and network. In recent years, meeting the requirements of DNN inference with limited hardware resources has become a challenge. 

Because of the limitation of hardware resources and the demands of application capabilities, the common assumption has been that end devices cannot realize a large amount of computations with reasonable latency and energy consumption. Thus, cloud computing has emerged as a solution to this problem, because it provides infinite computing  storage and resources to end-users based on their demands, anywhere and at any time~\citep{kumar2019comprehensive}.
However, cloud computing causes high latency and requires high transmission bandwidth. In addition, the cloud is usually unreliable~\citep{lin2017survey}. Edge computing has been proposed to compensate for these drawbacks. In the emerging edge computing~\citep{shi2016edge}, edge nodes are usually closer to the sensors than the remote cloud, resulting in the advantage of low transmission delay and the disadvantage of limited resources.
The advantages of cloud computing and edge computing have led to them being used in DNN-driven applications and piqued the interest of researchers.

DNNs run on the cloud because of a lack of processing capacity of end devices; this requires data transmission to the cloud through a wireless network, imposing significant computational pressure on the data center. In addition, running DNNs on the cloud may result in high latency and require high transmission bandwidth. In edge computing, the energy and accuracy of DNN-driven applications are limited because of resource constraints.
Therefore, DNN partition was proposed in recent years to split the DNN into several parts and offload them to the specified deployment locations.

DNN partition has made progress in various cognitive services~\citep{ding2020cognitive}. For example, DNN partition has been widely applied in wearable cameras used for recognizing objects and understanding the surrounding environment, because it can overcome the limitations of mobile devices and the unsatisfactory responses of these cameras.  
In smart healthcare and disease detection,  minimizing response latency and ensuring user experience are extremely importance~\citep{zeng2020coedge}. DNN partition has shown unprecedented ability in processing human-central contents, such as learning abstract representation and extracting high-level features; moreover, it has resolved the limited source in edge devices while protecting patients' privacy when offloading data in the cloud. Therefore, studying DNN partition is useful.  

DNN partition approaches over cloud, edge, and end devices have been investigated in many studies. However, a survey on the overall framework in DNN partition approaches is lacking. In this paper, we  systematically review the typical partition approaches to facilitate researchers' understanding.

\subsection{Contributions and Paper Organization}

In this study, we first acquire a comprehensive literature on DNN partition approaches using major search engines and digital libraries. Then, we systematically review the DNN partition approaches over cloud, edge, and end devices. Subsequently, we introduce the application scenarios and general definition of DNN partition. The main contributions of this study are listed as follows:

\begin{itemize}
\item We summarize the technical contributions of related studies and describe the five-dimensional classification framework for DNN partition approaches.
\item We propose metrics for evaluating and comparing different DNN partition approaches.
\item We highlight and discuss some challenges and present potential future research directions.
\end{itemize}

This paper is organized as follows. Section~\ref{sec:3} introduces application scenarios for DNN partition and the general mathematical definition of the DNN partition problem. The implementation technologies for offloading DNN partition models are also introduced. Section~\ref {sec:4} presents the classification framework, which consists of five factors for describing DNN partition approaches.
Section~\ref{sec:5} describes the refinement of the metrics of DNN partition models and the analysis and comparison of the typical partition approaches based on these metrics. Section~\ref {sec:6} provides a discussion on future challenges and opportunities. Section~\ref{sec:7} concludes this paper.

\section{Paper Collection Methodology}

As a general framework, we followed the guidelines described by Kitchenham and Charters~\cite{keele2007guidelines}to plan and conduct our survey. We classified the collection of the papers into four phases. 

\begin{itemize}
\item\textbf{Phase 1}: We collected papers by using typical search engines and digital libraries, including Google Scholar, IEEE Xplore, ACM Digital Library, SpringerLink, DBLP, and arXiv.
\item \textbf{Phase 2}: We conducted exact keyword searches on these search engines and digital libraries to collect papers related to DNN partition approaches over cloud, edge, and end devices. This resulted in more than 1800 papers. We used the following search terms: deep learning, DNN(s), deep neural network, partition, splitting, split, offloading, deployment, uploading, edge computing, MEC, cloud, device, collaborative, joint(ly), resource-efficient, energy-efficient and delay.

\item \textbf{Phase 3}: We created the search string based on the search aforementioned research terms by splitting the keywords into three buckets. Each bucket was represented as an ``OR" relation of keywords, whereas the complete search string was an ``AND" relation between the three buckets. We considered 133 of the papers found by applying the following search string:

\begin{center}
    (``deep\ learning'' $\lor$ ``DNN'' $\lor$ ``deep\ neural\  network'') \\
    $\land$
    (``edge\ computing'' $\lor$ ``MEC'' $\lor$ ``cloud'') \\
    $\land$ 
    (``partition'' $\lor$ ``split'' $\lor$ ``offloading'' $\lor$ ``joint'' 
    $\lor$ ``uploading'' $\lor$
    ``deployment'' $\lor$ ``collaborative'')
\end{center}

\item \textbf{Phase 4}: After careful analysis of the three buckets, we filtered for quality by using exclusion criteria. This required a manual analysis of large parts of each publication. We performed one level of snowballing and analyzed the references and research cited in each included paper. We applied the inclusion and exclusion criteria, ensuring that essential papers missed using our selection of search engines and terms were found. Finally, we considered 60 out of the 133 papers collected for this survey. Only papers published before December 2021 were considered for this survey.
\end{itemize}

\section{Problem Definition}
\label{sec:3}

DNN partition involves the splitting of the DNN at the granularity of layers or to finer granularity. Parts of the DNN are then offloaded on cloud, edge, and end devices to improve DNN inference performance.
This section introduces the scenarios of DNN-driven applications, to highlight the necessity of DNN partitioning. Then, implementation technologies are introduced based on application scenarios. Finally, the mathematical definition of DNN partition is provided.

\subsection{Application Scenarios}

\begin{figure*}[ht]
   \centering
    \subfigure[Fall detection in smart home]{
    \includegraphics[width=0.4\textwidth]{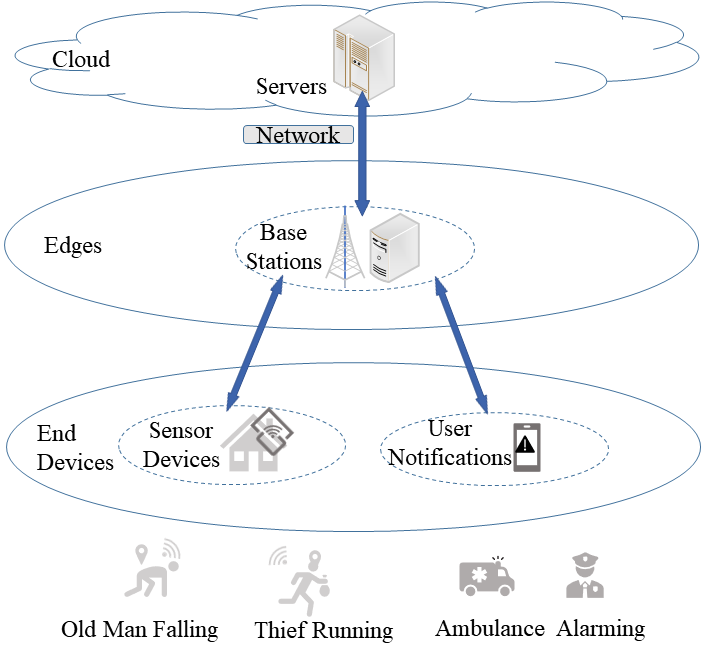}\label{Fall detection in smart home}
   }
   \subfigure[Smart traffic in smart city]{
    \includegraphics[width=0.4\textwidth]{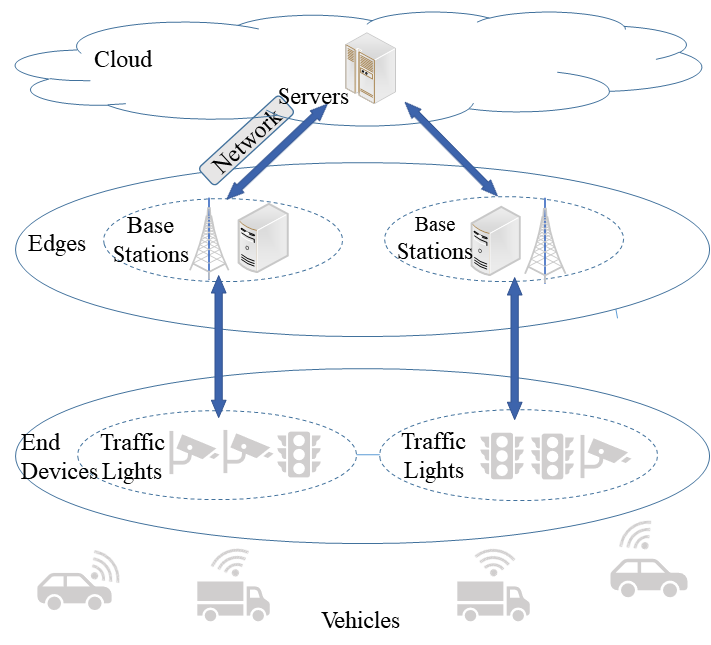}\label{Smart traffic in smart city}
   }
   \subfigure[Automatic control in smart industry]{
    \includegraphics[width=0.4\textwidth]{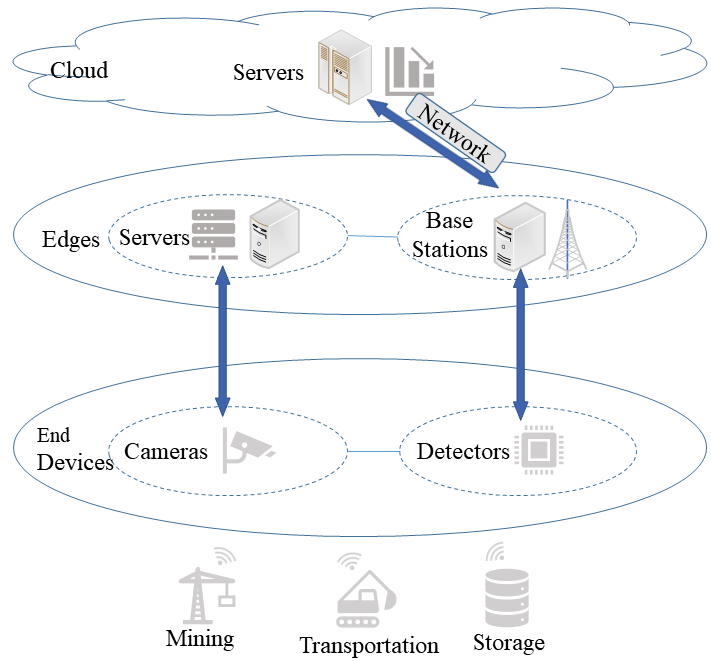}\label{Automatic mining in smart industry}
   }
    \subfigure[VR/AR]{
    \includegraphics[width=0.4\textwidth]{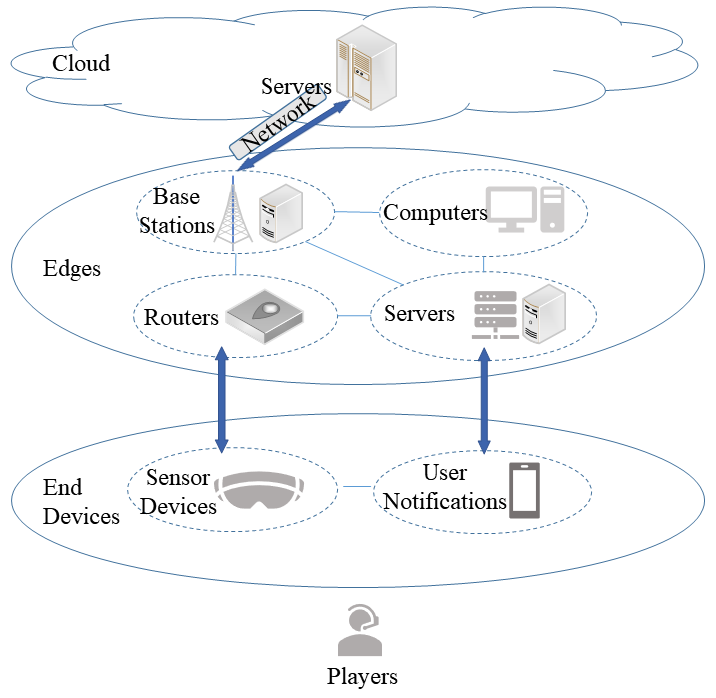}\label{VR/AR}
   }

   \caption{Application scenarios}\label{fig:scenario of traffic}
\end{figure*}

Owing to the rapid advancement in wireless communication technology (e.g., 5G, edge computing, and cloud), the number of IoT devices has increased dramatically, resulting in a massive amount of data. To fully utilize this data, deep learning has been widely adopted in many scenarios, such as smart cities, smart homes, and virtual/augmented reality (VR/AR), as shown in Fig.~\ref{fig:scenario of traffic}.
We introduce some typical DNN-driven applications in these scenarios below.

Smart cities have become a part of people's daily lives. For example, as shown in Fig.~\ref{Fall detection in smart home}, a smart home camera runs convolutional neural network (CNN)-based face recognition~\citep{GUNES20071334} to provide real-time inspection and warnings to protect the home.
Furthermore, the fall detection system~\citep{hsu2017fallcare+} generates an alert message when an object falls in the smart home.
DNNs are also widely used in smart traffic. For example, in Fig.~\ref{Smart traffic in smart city}, the edge video convergence node is connected to the local surveillance camera, providing AI capabilities to various stock cameras with different capabilities. 

Industrial parks deploy AI and digital analysis capabilities to achieve real-time industrial control intelligence in the edge and local devices. In Fig.~\ref{Automatic mining in smart industry}, a DNN achieves real-time processing and analysis of data and objects with characteristic values by deploying target recognition and mining surveillance capabilities to meet real-time monitoring requirements.

VR and AR are new technologies in various fields~\citep{schmoll2018demonstration}. For example, typical multiplayer games are designed to run in the cloud with all the gaming clients are connected to it, as shown in Fig.~\ref{VR/AR}.

The performance requirements are presented in the application scenarios of DNNs in the IoTs. For example, the response delay may last for a few seconds when running a local DNN because of the limited computing capacity. This delay may result in a poor user experience and a completely unusable service. Therefore, improving DNN performance using DNN partition technologies is critical.  

\subsection{Implementation Technologies}

Recent research has introduced the prototype of DNN partition approaches. Most DNN partition approaches are efficient in simulation environments; therefore, it is essential to illustrate how to deploy and run DNN partition over cloud, edge, and end devices.
Here, we introduce how DNN partition technologies are implemented. 

\begin{figure*}[ht]
\centering
\includegraphics[width=1.0\textwidth]{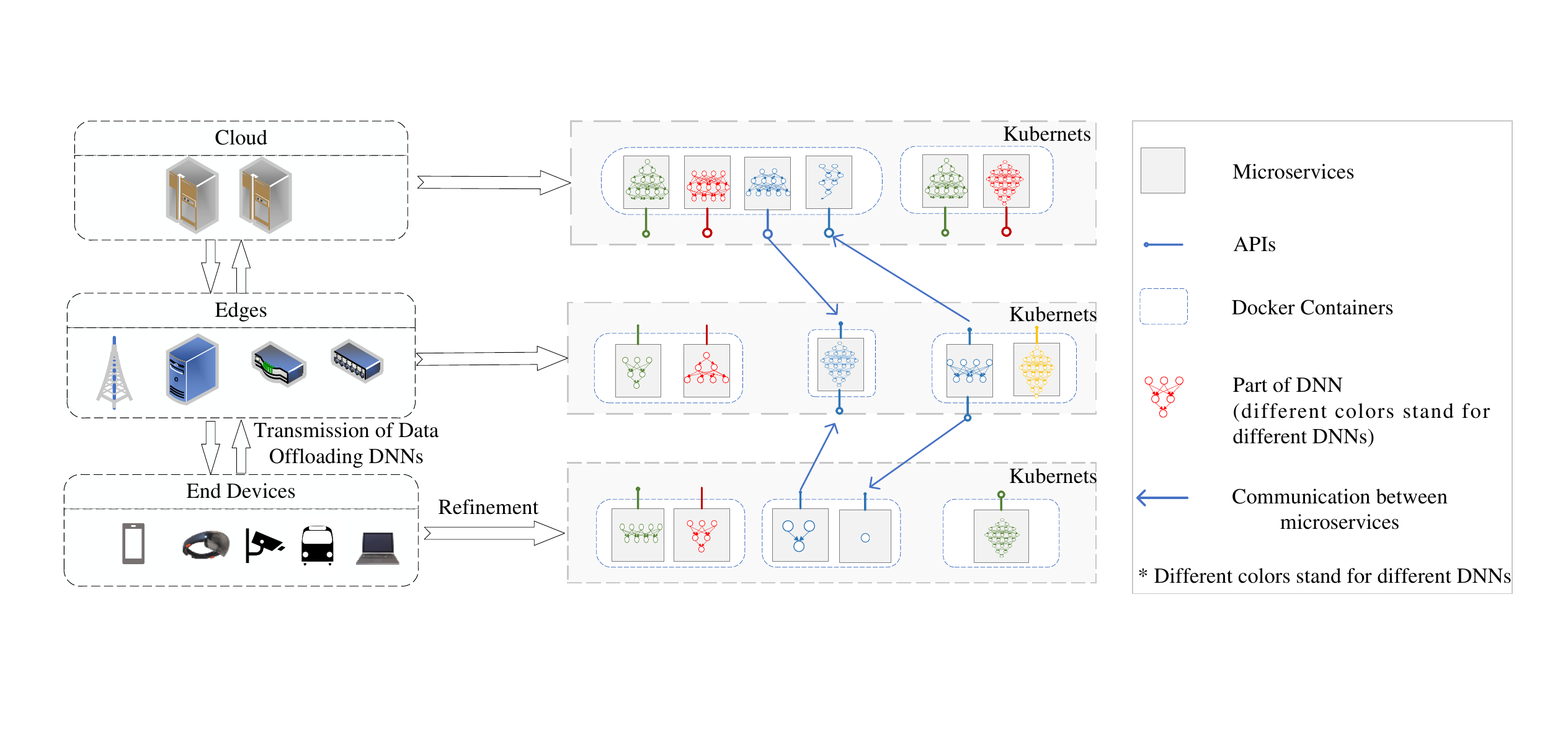}
\caption{\label{fig: deployment environment}Implementation technologies and runtime environment}
\end{figure*}

Each part of the DNN model is regarded as a microservice, and the container technology is adopted~\citep{kum2019deploying}. 
The DNN partition algorithm is offloaded in a master edge server at the container level~\citep{zhou2019distributing}, and micro-services with parts of the model are generated and packaged as containers.
Multiple microservices run an entire DNN model in containers across the end devices, edge devices, and cloud servers. This enables continuous delivery and deployment in large and complex services through API calls between microservices, which systems~\citep{balalaie2016microservices,satyanarayanan2017emergence}. In addition, tools such as Kubernetes, an open resource infrastructure for automated deployment and management of containerized applications~\citep{sayfan2017mastering, bernstein2014containers}, are employed to manage the containers.
The systems must be implemented at runtime. The overall framework is shown in Fig.~\ref{fig: deployment environment}.

\subsection{Mathematical Models}

This subsection details the definition of DNN partition over cloud, edge, and end devices and presents the mathematical model of this problem.

The output of a DNN partition model depends on the characteristics of the DNN, size of input data, memory footprint, battery, energy consumption, deployment location, network bandwidth, and number of deployed devices.
Therefore, we formulate an ordinary DNN partition model based on these factors.

\noindent
\textbf{Definition 1} (Communication). The communication among the cloud, edge, and end devices can be modeled by a graph $G_1\cup G_2\cup G_3 ={(D,E_{d,e})\cup (D,E_{d,c})\cup(M,E_{e,c})}$, where 
\begin{itemize}
    \item $D = \{1,2,\dots,|D|\}$ denotes the set of end devices. $M = \{1,2,\dots,|M|\}$ denotes the set of edges.
    \item $E_{d,e} = \{1,2,\dots,|E_{d,e}|\} $ denotes the set of physical links connecting edges to the end devices, $E_{d,c} = \{1,2,\dots,|E_{d,c}|\} $ denotes the set of physical links connecting cloud to the end devices, $ E_{e,c} = \{1,2,\dots,|E_e|\}$ denotes the set of physical links connecting cloud to the edges, and $B^w_{n,m}$ denotes the bandwidth of wireless link between the devices.
    \item $S=\{ S_1,S_2,\dots, S_N\}$ defines the set of the size of the input data of DNN, where the number of DNNs is denoted by $N$.
\end{itemize}

\noindent
\textbf{Definition 2} (Partition strategy). A DNN partition strategy can be defined as the set of several DNN partition results denoted by $O = \{O_{1},O_{2},\dots,O_{l}\}$. The type-n DNN partition result is denoted by $O_n = \{O_{n,1},O_{n,2},\dots,O_{n,l}\}$, where $O_{n,l}$ denotes a partition point.

\noindent 
\textbf{Definition 3} (Performance). The type-n indicators are defined as $P^n = \{P^n_1,P^n_2,\dots,P^n_\alpha\}$, which denote the accuracy of DNN inference, delay time of DNN inference,  energy consumption and others. Furthermore, $p^r_{n,a} = \{p^{r,a}_{n,0},p^{r,a}_{n,1},\dots,p^{r,a}_{n,k}\}$, $p^{l,a}_n = \{p^{l,a}_{n,0},p^{l,a}_{n,1},\dots,p^{l,a}_{n,k}\}$ and $p^{f,a}_n = \{p^{f,a}_{n,0},p^{f,a}_{n,1},\dots,p^{f,a}_{n,k}\}$ denote the sets of type-$a$ performance for type-n DNN inference on the cloud, end and edge device, respectively, and $p^{t,a}_n = \{p^{t,a}_{n,0},p^{t,a}_{n,1},\dots, p^{t,a}_{n,k}\}$ denotes the set of type-$a$ performance for DNN intermediate data transmission. Then, $ \forall a \in \alpha$, type-$a$ performance of type-n DNN is denoted as follows:
\begin{equation}
\label{E1}
P^{n}_{a} = \bigcup_{i=1}^{k}\{p^{r,a}_{n,i}\lor p^{f,a}_{n,i} \lor p^{l,a}_{n,i} \lor p^{t,a}_{n,i}\}
\end{equation}
In addition, we define a representative function as follows: 

\begin{subequations}
\begin{numcases}{\sigma(i,u) =}
1, & \textrm{$i \in u$}\\
0, & \textrm{$i \not\in u$}
\end{numcases}
\end{subequations}
where $i$ indicates whether or not layer $i$ of the DNN is on device $u$. 

\noindent
\textbf{Modeling}. The optimal objective of partitioning and offloading over cloud, edge, and end devices is to improve the DNN inference performance.
The partition model of the total optimal performance , (also the optimization target) on a distributed system can be formulated as follows:

\begin{equation}
\setcounter{equation}{3}
\label{E3}
\begin{split}
P^a_{total} = \max/\min\{ &\sum_{n=0}^N\sum_{i=0}^{k_n}\{p^{r,a}_{n,i}\sigma(i,r)+p^{f,a}_{n,i}\sigma(i,f)+p^{l,a}_{n,i}\sigma(i,l)\\
&+p^{t,a}_{n,i}\prod_{j\in\{r,f,l\}}[\sigma(i,j)(1-\sigma(i+1,j))]\}\}.
\end{split}
\end{equation}

s.t. 
\begin{equation}
\label{E4}
\sigma(i,r)+ \sigma(i,f)+\sigma(i,t) = 1,   \forall i\leq N
\end{equation}

\begin{equation}
\label{E5}
\begin{split}
\bigvee^{\alpha}_{b\neq a,b = 1}\{\sum_{n=0}^N\sum_{i=0}^{k_n}\{p^{r,b}_{n,i}\sigma(i,r)+p^{f,b}_{n,i}\sigma(i,f)+p^{l,b}_{n,i}\sigma(i,l)\\
+p^{t,b}_{n,i}\prod_{j\in\{r,f,l\}}[\sigma(i,j)(1-\sigma(i+1,j))]\}\leq \hat{C_b}\}.
\end{split}
\end{equation}
where $\hat{C_b}$ indicates the constraint boundary of performance $b$. Note that there can be multiple optimal objectives because Eq.~\ref{E3} comprises multiple performance indicators. 

\section{Classification Framework}
 \label{sec:4}
 
\subsection{Overview}

\begin{figure*}[ht]
\centering
\includegraphics[width=1\textwidth]{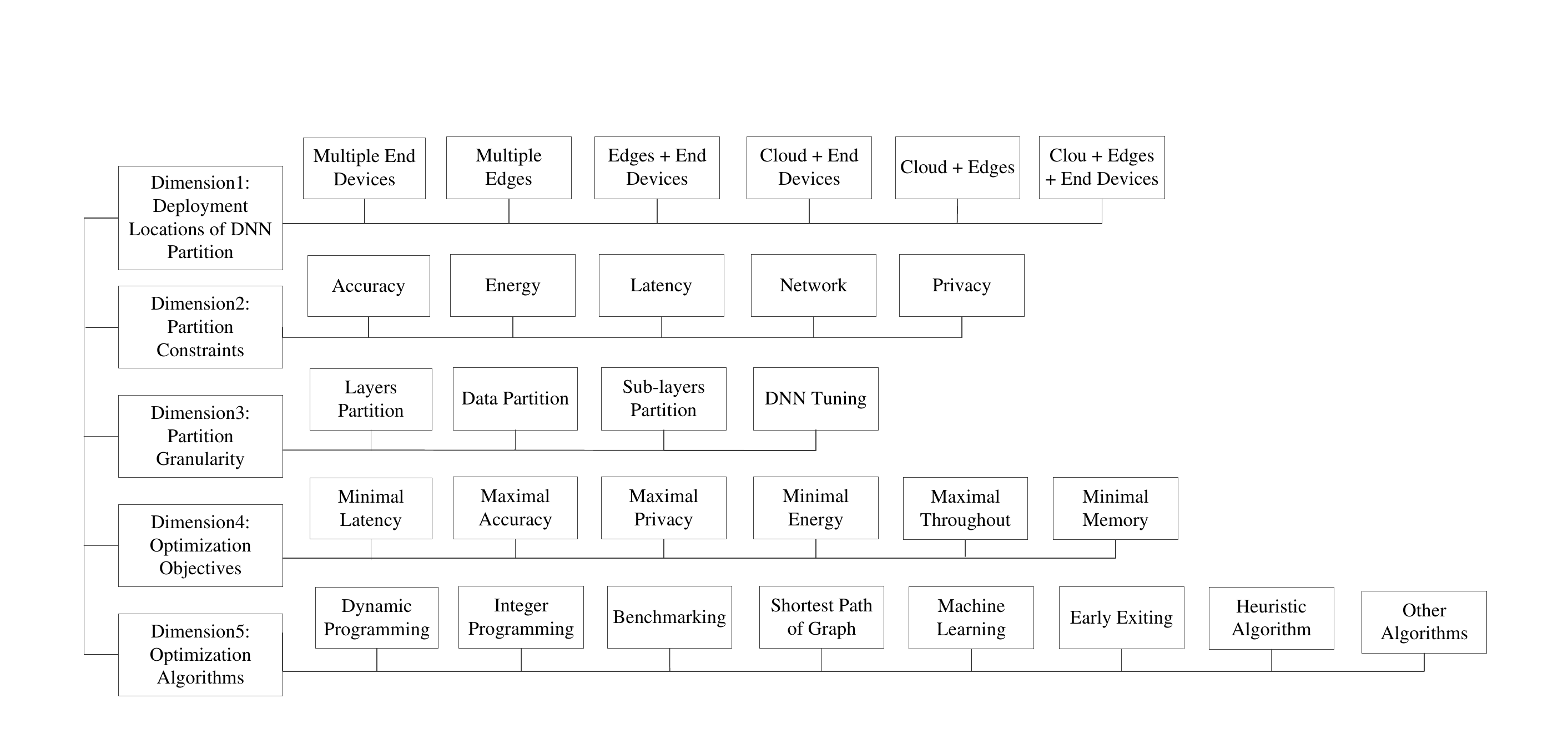}
\caption{\label{fig:framework4.1}Classification framework}
\end{figure*}

A five-dimensional classification framework for DNN partition approaches is described. First, we answer why, how, and what DNN partition is. Because of the limited resources and performance requirements, the DNN is divided into some parts; thus, the constraint and optimization objectives are two main factors in DNN partition. Then we seek to understand how to divide the DNN model partitions; this depends on the partition granularity: the layers, sub-layers, and input data. Therefore, this is one of the main factors influencing the DNN partition strategy.
Furthermore, the DNN partition problem is formulated, we decide the optimal partitions, obtained by the optimization algorithms, which comprise machine learning, dynamical programming, and others. Finally, the parts of DNN are offloaded to the deployment locations, which are usually provided in advance.
Based on these five factors, the systematic classification framework of the DNN partition is shown in Fig.~\ref{fig:framework4.1}. In addition, the DNN partition approaches employed in recent studies are classified in this framework, 
as shown in Table~\ref{tab:DNN}.

\begin{table}[ht]
\scriptsize
\centering
\caption{\label{tab:DNN} Reference descriptions in classification framework}
\begin{tabularx}{\textwidth}
{
  >{\hsize=.15\hsize\linewidth=\hsize }X
  | >{\hsize=.2\hsize\linewidth=\hsize \raggedright\arraybackslash}X
  | >{\hsize=1.65\hsize\linewidth=\hsize}X
}
\hline
\bf{Framework}&\multicolumn{1}{c}{\bf Description}&\multicolumn{1}{c}{\bf Reference}\\\hline
\multirow{6}{*}{\makecell[l]{Deployment \\ Locations}} & Multiple End Devices &\citep{mao2017mednn, zhou2019distributing, guo2021dynamic}\\\cline{2-3}
& Multiple Edges &\citep{miao2020adaptive, zeng2020coedge, he2020joint, xue2020edgeld, zhao2018deepthings}\\\cline{2-3}
& Edges + End Devices &\citep{jeong2018ionn, li2018edge, ali2019deep, tang2020joint, xu2020energy, tian2021mobility, ren2020edge, mohammed2020distributed, li2021dnn, shi2019privacy, shin2019enhanced, ren2021fine, yang2021cooperative, jeong2020perdnn, wang2019adda}\\\cline{2-3}
& Cloud + Devices &\citep{kang2017neurosurgeon, han2016mcdnn, eshratifar2019jointdnn, li2018jalad, duan2021joint, xia2019dnntune}\\\cline{2-3}
& Cloud + Edges &\citep{ding2020cognitive, ko2018edge, fang2019serving, hu2019dynamic, gao2021pripro, li2021appealnet, mudassar2018edge, ding2020cloud}\\\cline{2-3}
& Cloud + Edge + End Devices & \citep{lockhart2020scission, xu2020collaborative, teerapittayanon2017distributed, ashouri2020analyzing, huang2020deepadapter, chen2021energy, hu2021enable, huang2019cost, zhang2020towards, ren2021fine, zhang2021towards, lin2019distributed}\\\hline
\multirow{4}{*}{\makecell[l]{Partition \\ Granularity}}& DNN Tuning&\citep{teerapittayanon2017distributed, han2016mcdnn, huang2020deepadapter, ding2020cognitive, li2018edge, li2018jalad, ko2018edge, li2021appealnet, hu2021enable, li2021dnn, mao2017mednn, xue2020edgeld, ding2020cloud}\\\cline{2-3}
& Data Partition & \citep{li2021appealnet, zeng2020coedge, mao2017mednn}\\\cline{2-3}
& \multirow{1}{*}{Layers Partition} &\citep{kang2017neurosurgeon, lockhart2020scission, teerapittayanon2017distributed, huang2019cost, jeong2018ionn, eshratifar2019jointdnn, li2018edge, li2018jalad, ko2018edge, fang2019serving, chen2021energy, hu2019dynamic, tang2020joint, xu2020energy, hu2021enable, tian2021mobility, ren2020edge, duan2021joint, he2020joint, li2021dnn, mao2017mednn, shi2019privacy, yang2021cooperative, xue2020edgeld, ren2021fine, zhang2021towards, ding2020cloud, xu2020collaborative, jeong2020perdnn, lin2019distributed, wang2019adda, guo2021dynamic}\\\cline{2-3}
& Sub-layers Partition & \citep{miao2020adaptive, mohammed2020distributed, yang2021cooperative, zhao2018deepthings}\\\hline

\multirow{5}{*}{\makecell[l]{Partition \\ Constraints}}&\multirow{1}{*}{Energy}&\citep{miao2020adaptive, lockhart2020scission, teerapittayanon2017distributed, ashouri2020analyzing, huang2019cost, han2016mcdnn, huang2020deepadapter, eshratifar2019jointdnn, ali2019deep, tang2020joint, xu2020energy, hu2021enable, zeng2020coedge,he2020joint, mohammed2020distributed, shi2019privacy, zhou2019distributing, xu2020collaborative}\\\cline{2-3}
&Utilities&\citep{lockhart2020scission, ashouri2020analyzing, huang2020deepadapter, eshratifar2019jointdnn, hu2019dynamic, ali2019deep, mohammed2020distributed, xue2020edgeld, ren2021fine, zhou2019distributing, xu2020collaborative, zhang2021towards, jeong2020perdnn, guo2021dynamic}\\\cline{2-3}
&Accuracy&\citep{teerapittayanon2017distributed, han2016mcdnn, li2018jalad, hu2021enable, tian2021mobility, wang2019adda}\\\cline{2-3}
&Delay&\citep{ashouri2020analyzing, han2016mcdnn, huang2020deepadapter, ding2020cognitive, eshratifar2019jointdnn, chen2021energy, hu2019dynamic, ali2019deep, xu2020energy, li2021appealnet, zeng2020coedge, huang2019cost, li2021dnn, xu2020collaborative, lin2019distributed}\\\cline{2-3}
&Privacy&\citep{gao2021pripro}\\\hline
\multirow{7}{*}{\makecell[l]{Optimization \\ Objectives }}&\multirow{1}{*}{Delay}&\citep{kang2017neurosurgeon, miao2020adaptive, lockhart2020scission, teerapittayanon2017distributed, ashouri2020analyzing, jeong2018ionn, eshratifar2019jointdnn, li2018edge, li2018jalad, hu2019dynamic, tang2020joint, hu2021enable, tian2021mobility, ren2020edge, duan2021joint, he2020joint, mohammed2020distributed, li2021dnn, mao2017mednn, shi2019privacy, yang2021cooperative, xue2020edgeld, ren2021fine, ding2020cloud, jeong2020perdnn, wang2019adda, guo2021dynamic}\\\cline{2-3}
&Throughput &\citep{lockhart2020scission, ko2018edge, hu2019dynamic, zhang2021towards, lin2019distributed}\\\cline{2-3}
&Utilities  &\citep{xu2020energy}\\\cline{2-3}
& Energy &\citep{lockhart2020scission, ashouri2020analyzing, huang2019cost, eshratifar2019jointdnn, ko2018edge, chen2021energy, ali2019deep, xu2020energy, zeng2020coedge, ren2021fine, zhang2021towards}\\\cline{2-3}
&Privacy  &\citep{shi2019privacy}\\\cline{2-3}
&Accuracy&\citep{han2016mcdnn, huang2020deepadapter, ding2020cognitive, gao2021pripro, li2021appealnet, li2021dnn, xu2020collaborative}\\\cline{2-3}
&Memory&\citep{zhao2018deepthings}\\\hline

\multirow{11}{*}{\makecell[l]{Optimization \\ Algorithms}} & Dynamic Programming&\citep{li2021dnn}, \citep{zhou2019distributing},\citep{zhang2021towards, lin2019distributed}\\\cline{2-3}
&Integer Programming&\citep{han2016mcdnn, eshratifar2019jointdnn, chen2021energy, tang2020joint, xu2020energy, hu2021enable, he2020joint, xu2020collaborative}\\\cline{2-3}
&Benchmarking&\citep{lockhart2020scission,ashouri2020analyzing, huang2020deepadapter, ren2020edge, xia2019dnntune}\\\cline{2-3}
&Shortest Path of Graph&\citep{jeong2018ionn, eshratifar2019jointdnn, zeng2020coedge, jeong2020perdnn}\\\cline{2-3}
&Machine Learning&\citep{xu2020energy, li2021appealnet, ren2021fine}\\\cline{2-3}
&Early Exiting&\citep{teerapittayanon2017distributed, li2018edge, hu2021enable, ren2020edge, li2021dnn, wang2019adda}\\\cline{2-3}
&Markov Approximation&\citep{he2020joint}\\\cline{2-3}
&Game Approach&\citep{mohammed2020distributed}\\\cline{2-3}
&Convex Optimization&\citep{li2021dnn}\\\cline{2-3}
&Lyapunov Optimization&\citep{shi2019privacy, xu2020collaborative}\\\cline{2-3}
&Heuristic Algorithms &\citep{miao2020adaptive, huang2019cost, jeong2018ionn, han2016mcdnn, huang2020deepadapter, ding2020cognitive, chen2021energy, hu2019dynamic, tang2020joint, xu2020energy, tian2021mobility, li2021dnn, mao2017mednn, wang2019adda, guo2021dynamic}\\\hline
\end{tabularx}
\end{table}

\subsection{Dimension 1: Deployment Locations of DNN Partition}

We first describe DNN partition's deployment locations over cloud, edge, and end devices. The structure of deployment locations after partitioning has six categories: distributed computing across the cloud + end devices, edges + end devices, cloud + edges,  cloud + edges + end devices, multiple end devices, and multiple edges. In this study, edge nodes included edge servers, base stations, and others. The devices that obtained the data were regarded as end devices.

\subsubsection{Cloud + End Devices}
 
The DNN partition approach is widely used on end devices and cloud collaboration.
In a DNN partition approach, some parts of the DNN are inferred locally, and others are offloaded to the cloud. Today, research has achieved collaborative inference technology between the cloud and single, multiple, or mobile end devices. 
Primarily, Neurosurgeon~\citep{kang2017neurosurgeon} can orchestrate the distribution of computation between mobile end devices and the cloud. Approximate model scheduling (MCDNN)~\citep{han2016mcdnn} and joint optimization of DNN Partition and scheduling~\citep{duan2021joint} are presented for multiple DNN inferences on a single cloud and multiple user devices.
JointDNN~\citep{eshratifar2019jointdnn} with AppealNet~\citep{li2021appealnet} is also a method to deploy the parts of DNN on the cloud and end devices.

\subsubsection{Edges + End Devices}
 
Similar to cloud and end devices, researchers have focused on on the collaboration between the edge and end devices. 
Initially, the DNN was only partitioned on one end device and one edge server~\citep{ali2019deep, li2018edge}. However, computation offloading must be jointly handled with resource allocation among users in the case of more general multi-users. Thus, joint multi-user DNN partitioning based on multilevel offloading is proposed~\citep{tang2020joint,li2021dnn}.
Furthermore, IONN~\citep{jeong2018ionn} and EPDNN~\citep{shin2019enhanced} consider the mobility of clients or the edge server for a single DNN. In addition, DNN partition approaches are processed between some edge servers and mobile end devices~\citep{tian2021mobility, ren2020edge, mohammed2020distributed,jeong2020perdnn}.

\subsubsection{Cloud + Edges}

To meet the users' demands for fast response, long duration, and high accuracy, cloud-edge collaboration computing is proposed. The end device does not require the capability to compute because it only sends task requests and data to edge servers~\citep{ding2020cognitive, kum2019deploying, ko2018edge,fang2019serving, hu2019dynamic, gao2021pripro}. Furthermore, some researchers have considered the structure of DNN, such as chain-DNN and DAG-DNN, when they split the DNN into parts on edge devices or the cloud~\citep{fang2019serving, hu2019dynamic}.

\subsubsection{Cloud + Edges + End Devices}

The DNN partition over cloud, edge, and end devices has been an important research topic in recent years because it not only considers the characteristics of the end device and edge, but also the properties of the cloud~\citep{ashouri2020analyzing, lockhart2020scission, huang2019cost,teerapittayanon2017distributed}. 
In DDNN~\citep{teerapittayanon2017distributed}, one part of the DNN runs on a single device, and the intermediate DNN output is sent to the cloud.
Similarly, a collaborative framework has been presented that connects the mobile web to edge and remote cloud servers~\citep{huang2020deepadapter}.  Furthermore, one DNN partition~\citep{hu2021enable} has presented a pipeline execution model for the mobility of devices and multiple DNNs.
\subsubsection{Multiple End Devices}

The multiple end devices synergy has also been considered. MeDNN~\citep{mao2017mednn} is a local distributed mobile computing system with enhanced partitioning and deployment, which is tailored for large DNNs. The parallel processing of DNN inference across multiple heterogeneous devices is still being explored~\citep{zhou2019distributing}.
\subsubsection{Multiple Edges}
In addition, the deployment locations include multiple end devices.
CoEdge~\citep{zeng2020coedge} orchestrates a single DNN inference over heterogeneous edge devices. Furthermore, multiple DNNs partition inference in MEC and DNN training on the cloud, which is analyzed to accurately minimize the End-to-End (E2E) delay~\citep{he2020joint}.

\subsection{Dimension 2: Partition Granularity}

This subsection describes how to partition DNN models. One typical method is to partition the DNN into layers or sub-layers. The other is to tune the DNN model.

\subsubsection{DNN Partition}

The DNN is split into parts at the granularity of layers, input data, or sub-layers.

\textbf{Layer Partition}.
In general, the DNN is split into two parts. Neurosurgeon~\citep{kang2017neurosurgeon}, Edgent~\citep{li2018edge}, JALAD~\citep{li2018jalad}, and PriPro~\citep{gao2021pripro} are typical partition methods that split a DNN into two parts. Some DNN partition approaches~\citep{ko2018edge, hu2019dynamic, li2021dnn} are similar in that the output of the partition is only two parts.
In addition, the DNN can be subdivided into two more parts at the granularity of layers. 
JointDNN~\citep{eshratifar2019jointdnn} is a new method that allows computation on either platform for each layer independently of the other layers; this may allow one more partition point across the mobile device and cloud. However, compared with the general solutions, the approach in~\citep{ren2021fine} is more flexible with regard to the fine-grained DNN computation partitioning mechanism.
The number of DNNs influences DNN partition. The joint optimization of multiple DNN partition and scheduling for mobile cloud~\citep{duan2021joint} splits each DNN into two parts at the granularity of each DNN layer and the different partition points. Moreover, to address the multiple DNN partition problem, a DNN partition strategy with layer partition operations is also considered to be efficient~\citep{chen2021energy,hu2021enable}. 

\textbf{Data Partition}.
DNN partition can be employed at the granularity of the input data. The input data are split and processed on multiple devices simultaneously at runtime. For instance, CoEdge~\citep{zeng2020coedge} divides the input data and reserves model parameters for the given DNN model.  
AppealNet~\citep{li2021appealnet} is a unique method that joints edge devices and the cloud according to the complexity of the input data; the light DNN is deployed in the devices, and the heavy DNN is uploaded to the cloud.

\textbf{Sub-layer Partition}. In addition to the aforementioned DNN partitions, a DNN partition at a finer granularity has also been proposed.
An adaptive DNN partition algorithm is presented at the granularity of branches in each layer~\citep{miao2020adaptive}. Mohammed et al. ~\citep{mohammed2020distributed} proposed a fine-grained adaptive partitioning method to split a DNN into pieces that can be smaller than a single layer. All parallel paths in the DNN are considered, depending on the convolutional or fully-connected layer type.
Homoplastically, there is a DNN partition to slice the original CNN layer stacks into independently distributed execution units, and each unit occupies a small memory footprint~\citep{zhao2018deepthings}.

The multi-granularity of the DNN partition is considered because a DNN partition model that combines two or more partition granularities can improve the DNN performance. For example, Yang et al.~\citep{yang2021cooperative} leveraged the data and layer partition by dividing a DNN model into several blocks and processing each block differently. Furthermore, this method splits the input data of each layer to divide the computation in a block into independent tasks performed by different edge devices.

\subsection{Dimension 3: Partition Constraints}

The DNN partition is an optimization problem that addresses the limited resources and user requirements. Therefore, the partition constraints become one of the most indispensable factors that influence the DNN partition model.

Generally, the DNN partition relates to deployment resources, such as the deployment location, the limitation of device and edge servers, network bandwidth, and the model's property. Naturally, the constraint also contains user requirements; for example, the inference accuracy cannot be lower than that for the user requirements. 
We introduce the constraints from two aspects in the following section.

\subsubsection{DNN Tuning}

One method deserves to be mentioned is the fine-tuning of DNN usually used in DNN partitions. The purpose of tuning DNN is to overcome two challenges. One is decreasing the DNN parameters' redundancy to within the required accuracy, such as pruning; a lightweight DNN can optimize resource utilization. The other challenge is to decrease the relevance between DNN layers, such as layer fusion. When the output of one layer is the input of other layers, large quantities of data result in transmission delay. 
Therefore, tuning DNN is the most common method to improve the performance of DNN-driven applications. Fine-tuning DNN is achieved by tuning the internal parameters or structure to meet the requirements of applications. For example, fine-tuning methods include parameters binarization, matrix factorization, pruning, compression, and others. The main tuning methods in recent papers are described in Table~\ref{tab:tunes}.

\begin{table*}[ht]
\footnotesize
\centering
\caption{\label{tab:tunes} Studies in which different tuning methods were adopted.}
\begin{tabular}{l|l}\hline
\multicolumn{1}{c|}{\bf Reference}&\multicolumn{1}{c}{\bf Tuning Methods}  \\\hline
BranchyNet~\citep{teerapittayanon2017distributed}&
Parameters Binarization \\\hline
\multirow{2}{*}{MCDNN~\citep{han2016mcdnn}} & Matrix Factorization, Pruning\\
&Architectural change\\\hline
DeepAdapter~\citep{huang2020deepadapter} & Pruning\\\hline
MeDNN~\citep{mao2017mednn} & Pruning, Quantization Compression\\\hline
AppealNet~\citep{li2021appealnet} & Architectural change\\\hline
CSMEC~\citep{ding2020cognitive} & Parameter Sharing \\\hline
Edgent~\citep{li2018edge} & DNN Right-sizing  \\\hline
JALAD~\citep{li2018jalad} & Quantization Compression\\\hline
PADCS~\citep{hu2021enable} & Quantization Compression\\\hline
Edge-host Partitiong~\citep{ko2018edge} & Encoding Compression \\\hline
EdgeLD~\citep{xue2020edgeld} & Layer-fusion\\\hline
ShadowTutor~\citep{chung2020shadowtutor} & Knowledge Distillation\\\hline
\end{tabular}

\end{table*}

\subsubsection{Resource Constraints}

The aim in most studies was to enable DNN computation on resource-constrained mobile devices by partitioning DNNs horizontally or vertically into different sub-networks.
The limited memory of end devices and communication costs are the primary constraints. The BranchyNet~\citep{teerapittayanon2017distributed} is a typical algorithm that balances the accuracy and resource constraints. In addition, energy consumption is a type of DNN partition constraint~\citep{huang2019cost, gao2021pripro}. 
DeepAdapter~\citep{huang2020deepadapter} is a new DNN partition approach involving the network bandwidth constraint. The constraints dynamically consider the network's changes~\citep{hu2019dynamic}. Furthermore, a DNN partition strategy is constrained because the bandwidth allocated to all mobile devices covered by base stations cannot exceed the total bandwidth~\citep{he2020joint}. 

Some DNN partition approaches consider more comprehensive constraints. For example, the collaborative constraints of memory size, device energy budget, and cloud-cost budget are considered~\citep{han2016mcdnn,lockhart2020scission,li2021dnn}. 
Moreover, the parameters of the real-time adaptive model~\citep{eshratifar2019jointdnn} depend on the following factors: mobile and cloud hardware and software resources, battery capacity, network specifications, and inference delay requirement.
A new DNN partition method~\citep{xu2020energy} considers the energy efficiency of both user devices and base stations in a 5G-enabled MEC along with the devices and base stations' capabilities, the number of DNN tasks, and latency constraint.

\subsubsection{Self-Model Constraints}

When partitioning the DNN, we often need to compromise some performance to meet other requirements, which is unrealistic regardless of performance loss. Therefore, the constraints of DNN performance have been considered in recent studies. Constrained by the performance of DNNs, Chen et al.~\citep{chen2021energy} and Zeng et al.~\citep{zeng2020coedge} considered the latency requirements. JALAD~\citep{li2018jalad} considered the minimum accuracy requirement and number of DNN partitions.
Encoding feature spacing on the intermediate layers~\citep{ko2018edge} constrains accuracy as do data compression and early exiting algorithms~\citep{hu2021enable}.

\subsubsection{Privacy}

Privacy protection is essential in cloud-edge collaboration. Privacy protection is often compromised when considering the load on an edge device~\citep{osia2018deep}. DNN partition must also alert to the privacy issue. Sending intermediate DNN data from edge devices to the cloud is at risk of interception during various stages; therefore, PriPro~\citep{gao2021pripro} was introduced to protect privacy.

\subsection{Dimension 4: Optimization Objectives}

DNN-driven applications have different optimization requirements, such as the lowest latency to obtain inferred results and the client's minimum energy consumption. 
We now summarize the optimization targets into six categories based on relevant research in recent years. First, minimization latency is the most studied topic in relevant references.
Minimizing the overall delay of a frame is also an optimization objective~\citep{hu2019dynamic}.
An algorithm~\citep{miao2020adaptive} was proposed to balance multiple devices' loading rates and minimize latency. Edgent~\citep{li2018edge} was proposed as a solution to low latency edge intelligence.
The optimization objective for the multi-DNNs partition algorithm is to minimize the maximum DNN inference latency among all devices to reduce the global latency~\citep{tang2020joint}.
DeepAdapter~\citep{huang2020deepadapter} incorporates the mobile devices' latency, network condition, and computing capability.

Minimizing energy consumption or maximizing the accuracy of DNN inferences to achieve optimization objectives are also important. Reducing energy consumption decreases the cost of edge computing for offloading DNN-based applications to multiple DNNs~\citep{huang2019cost}. Moreover, the optimization computation schedule~\citep{eshratifar2019jointdnn} has been presented to meet the lowest energy consumption. Generally, 
multiple DNNs partition algorithms~\citep{chen2021energy} mainly aim to  minimize the energy consumption, with each DNN running an open loop, considering the runtime energy consumption per time unit and computing energy consumption.
Maximizing the accuracy is also regarded as an optimization objective~\citep{han2016mcdnn, ding2020cognitive}.
PriPro~\citep{gao2021pripro} injects noise for privacy protection targeting DNNs. Under this condition, the optimization objective is to maximize the accuracy. 

Generally, multi-objective optimization is more useful in practice than single-objective optimization. Scission~\citep{lockhart2020scission} can obtain an appropriate partition scheme according to the hardware conditions and user's demands, such as minimizing the latency and minimizing energy consumption. The optimal objectives of IONN~\citep{jeong2018ionn} not only reduce the latency but also consider the time to upload the DNN partitions.
DDNN~\citep{teerapittayanon2017distributed} also adopts layers partition to balance accuracy and energy consumption.
Furthermore, a trade-off study on the energy efficiency and throughput of the edge platform~\citep{ko2018edge} has been presented. 

\subsection{Dimension 5: Optimization Algorithms}

In different scenarios, many technologies employ different optimization objectives to obtain the optimal solution, including dynamic programming, integer programming (IP), convex optimization, reinforcement learning (RL), and the shortest path algorithm.
In particular, the constructed optimization objective function is generally NP-hard because of the non-linearity of the function and the uncertainty of the number of parameters. Therefore, researchers usually used approximate algorithms, including greedy algorithms, approximate convex optimization, and genetic algorithms.
In this section, we introduce some systematic algorithms.

\subsubsection{Dynamic Programming}

Dynamic programming is a classic algorithm used to solve the optimization problem.
The local solutions that are likely to be optimal are retained through decision making, and the others are discarded. Each subproblem is solved in turn, with the last sub-problem being the solution to the original problem~\citep{stuckey2020dynamic}. 
Dynamic programming is employed to obtain a set of optimal partition points for all devices algorithm~\citep{li2021dnn}, which minimizes the sum of total local computing time and the computing time on the edge server. CooAI~\citep{yang2021cooperative} adopts multi-layer partition and slicing (MLS) to solve the DNN inference optimal problem. MLS leverages dynamic programming by first computing and recording the optimal solution to each smaller subproblem and then reusing these solutions to solve a larger subproblem iteratively.

\subsubsection{Integer Programming}

IP is a subset of linear programming that only differs from linear programming in that it includes integer constraints. However, it is prevalently used to solve the optimization problem because of the mathematical definition of generalization. 
IP algorithms solve many DNN partition problems. For example, a nonlinear integer optimization problem can be formulated as an optimization partition problem~\citep{tang2020joint}. Furthermore, JALAD~\citep{li2018jalad} is formulated as an integer linear programming (ILP) problem. An exact solution is obtained by formulating an ILP for offloading with a single request~\citep{xu2020energy}. 
A joint optimization model of partition and resource allocation has been developed by establishing mixed-integer nonlinear programming~\citep{he2020joint}. 

\subsubsection{Benchmarking}

Benchmarking is similar to the listing technique and has strong universality. In this method, all the partition solutions are listed. However, the cost is high if the problem is complex or the number of solutions is massive. In some scenarios, there is no single optimal objective because the user's requirements change dynamically. Benchmarking is a standard method for specific static DNN partitions.
For example, the multiple criteria decision-making method based on the analytical hierarchy process (AHP) ~\citep{ashouri2020analyzing} adopts benchmarking to provide a DNN partition strategy.
Scission~\citep{lockhart2020scission} is a six-step methodology for automated partitioning. To find the valid partition points, Scission benchmarks each layer and block on the target hardware
resources and creates partition configurations from benchmark data.

\subsubsection{Shortest Path of a Graph}

The DNN inference process can be considered as a path from the beginning to the end of a graph. Each node of the DNN has several choices to be deployed in any environment (end device, edge device, server, etc); thus, it is a multipath from one node to another. Optimizing metrics such as delay and energy consumption can be regarded as the weight of the edge; therefore, the goal is to find the shortest path within the DAG formed by all the DNN computing layers.
A mobility-included DNN partition offloading algorithm (MDPO)~\citep{tian2021mobility} uses the shortest path to solve the optimization partition problem. 
In addition, the latency of uploading the DNN layer onto the server is considered. To decide optimal DNN partitions and uploading orders, IONN~\citep{jeong2018ionn} uses a novel graph-based algorithm and PerDNN~\citep{jeong2020perdnn} uses the shortest path of a DAG to partition and offload, considering the mobility of end devices.

\subsubsection{Machine Learning}

Machine learning is a promising method to handle high complexity. An optimal DNN partition decision for machine learning is determined according to the system’s state. It obtains this optimization decision using learning techniques. 
The RL online algorithm~\citep{xu2020energy} decides whether to wait for a subsequent request or select a request from the current arrival list. The reward function is the inverse of the average delay experienced by the admitted requests. Similarly, FEPD~\citep{ren2021fine} also adopts the RL algorithm.
Using the difference in the DNN partition, AppealNet~\citep{li2021appealnet} presents a two-head network architecture that consists of an approximator head, predictor head, and feature extractor.

\subsubsection{Early Exiting}

The early exiting mechanism is also proposed to improve the DNN inference. The main objective of early exiting is to terminate the inference process in an intermediate layer. The early exiting mechanism can avoid the forward process of the entire DNN through the input layer to the final layer. The existing early exiting methods include two categories~\citep{teerapittayanon2016branchynet}. The first category modifies added exit branches at specific layer locations in the standard DNN model structure and then trains the original model with the exit branches together. However, it is hard to find an exit layer for a given DNN, and an additional cost may occur due to the retraining model. The second category determines the exit point after the convolutional layer~\citep{panda2016conditional} before adding a classifier to determine whether the inference result is correct. In the research on DNN inference optimization, many methods combine DNN partition with an early exiting mechanism to enhance the DNN inference performance.
For example, the aggregation scheme ~\citep{teerapittayanon2017distributed}, Edgent~\citep{li2018edge}, ADDA~\citep{wang2019adda}, and offloading strategy optimization~\citep{pacheco2021calibration} are the DNN partition approaches that use early exiting mechanisms.

\subsubsection{Heuristic Algorithms}

Heuristic algorithms are proposed concerning the optimization algorithm.
The objective is to choose an efficient heuristic algorithm and obtain the best or sub-best solution. Some typical heuristic algorithms have solved optimization DNN partition problems in recent years.
An adaptive DNN partition algorithm~\citep{miao2020adaptive} is a type of heuristic algorithm. 
In addition, a discrete particle swarm optimization with genetic operators (DPSO-GO)~\citep{huang2019cost} has been used to find an offloading strategy by solving the NP-Hard problem to address the optimization problem. Similarly, a threshold-based workload partition algorithm~\citep{zeng2020coedge}, iterative alternating optimization algorithm (IAO)~\citep{tang2020joint}, greedy two dimensional partition (GTDP)~\citep{mao2017mednn}, and a binary-search-based partition algorithm~\citep{duan2021joint} were proposed to meet the NP-Hard problem. 

\subsection{Instantiation of Framework}

In summary, we discussed the recent research on DNN partition algorithms in terms of five dimensions, further verifying the research dimension's completeness.
For example, we described Energy-Aware Inference Offloading for DNN-Driven Applications~\citep{xu2020energy} in the classification framework. The {\bf deployment locations} were mobile end devices and edges, and the DNN was divided into several sub-parts at the {\bf granularity of layers}. The {\bf optimization goal} was constructed to achieve the minimum energy consumption considering the latency and limited edge resources as the {\bf partition constraints}. Thus, the {\bf optimization algorithm} was IP, the Random Rounding Approximation algorithm, and RL.
Similarly to the aforementioned description, we listed several typical algorithms and introduced their characteristics in terms of five dimensions. The reason for choosing these algorithms is that they cover all the possible values in each dimension (see Table~\ref{tab:Exapmle}).

\begin{table*}[ht]
\footnotesize
\centering
\caption{\label{tab:Exapmle} Using the proposed classification framework to delineate existing DNN partition approaches.}

\begin{tabularx}{\linewidth}{l|lllll}\hline
\multicolumn{1}{c|}{\multirow{2}{*}{\bf Reference}} &  \bf Deployment & \bf Partition  & \bf Partition  & \bf Optimization   & \bf Optimization \\
&\bf Locations&\bf Granularity& \bf Constraints&\bf Objectives&\bf Algorithms\\\hline
\multirow{3}{*}{DCOSD2D \citep{guo2021dynamic}}&\multirow{3}{*}{Co-Ends} &\multirow{3}{*}{Layers}&Computation&\multirow{3}{*}{Latency} &Greedy\\
&&&resource&& Heuristic\\
&&&Bandwidth&& KM\\\hline
\multirow{3}{*}{MeDNN \citep{mao2017mednn}}    &\multirow{3}{*}{Co-Ends}& Layers&Energy&\multirow{3}{*}{Latency}&\multirow{3}{*}{Greedy} \\
&&Data&\multirow{2}{*}{Bandwidth}\\
&&Tuning\\\hline
\multirow{2}{*}{ADPMEC \citep{miao2020adaptive}} &\multirow{2}{*}{Co-Edges }&\multirow{2}{*}{Sub-layers}&Bandwith&\multirow{2}{*}{Latency}&\multirow{2}{*}{Traversal}\\
&&&Deivce quantity\\\hline
\multirow{3}{*}{EdgeLD \citep{xue2020edgeld}}&\multirow{3}{*}{Co-Edges} &\multirow{2}{*}{Sub-Layers}&Bandwidth &\multirow{3}{*}{Latency} & \multirow{3}{*}{Traversal}\\
&& \multirow{2}{*}{Tuning} & Computation &&\\
&&&resource&&\\\hline
\multirow{2}{*}{DeepThings \citep{zhao2018deepthings}}&\multirow{2}{*}{Co-Edges} &Sub-Layers&Computation&\multirow{2}{*}{Memory} & \multirow{2}{*}{Traversal}\\
&&Tuning&resource&&\\\hline
\multirow{2}{*}{JMDP \citep{tang2020joint}}&\multirow{2}{*}{Edges-Ends}& \multirow{2}{*}{Layers}&Computation&\multirow{2}{*}{Latency} &IP\\
&&&resource&&IAO\\\hline
\multirow{2}{*}{PANDA \citep{shi2019privacy}}& \multirow{2}{*}{Edges-Ends}&\multirow{2}{*}{Layers}&\multirow{2}{*}{Energy}&Privacy& \multirow{2}{*}{Lyapunov}\\
&&&&Latency\\\hline
\multirow{2}{*}{EPDNN \citep{shin2019enhanced}}&\multirow{2}{*}{Edges-Ends} &\multirow{2}{*}{Layers}&Computation&\multirow{2}{*}{Efficiency} &\multirow{2}{*}{Greedy}\\
&&&resource&&\\\hline
\multirow{3}{*}{CoopAI \citep{yang2021cooperative}}&\multirow{3}{*}{Edges-End} &\multirow{2}{*}{Layers}&Bandwidth &\multirow{3}{*}{Latency} & \multirow{3}{*}{DP}\\
&& \multirow{2}{*}{Data}  & Computation  &&\\
&&&resource&&\\\hline
\multirow{2}{*}{PerDNN \citep{jeong2020perdnn}}&\multirow{2}{*}{Edges-Ends} &\multirow{2}{*}{Layers}&Computation&\multirow{2}{*}{Latency} & \multirow{2}{*}{Shortest Path}\\ 
&&&resource&&\\\hline
\multirow{2}{*}{ADDA \citep{wang2019adda}}& \multirow{2}{*}{Edges-Ends}& \multirow{2}{*}{Layers}&\multirow{2}{*}{Accuracy} & \multirow{2}{*}{Latency} & Greedy\\
&&&&& Early Exiting\\\hline
\multirow{2}{*}{JALAD \citep{li2018jalad}}&\multirow{2}{*}{Cloud-Ends}&Layer&\multirow{2}{*}{Accuracy}&\multirow{2}{*}{Latency}&\multirow{2}{*}{ILP}\\
&&Tuning\\\hline
\multirow{2}{*}{AppealNet \citep{li2021appealnet}}&\multirow{2}{*}{Cloud-Edges } &Data &\multirow{2}{*}{Energy }&\multirow{2}{*}{Accuracy}&\multirow{2}{*}{ML}\\
&&Tuning\\\hline
\multirow{2}{*}{TREND-WANT \citep{lin2019distributed}}&\multirow{2}{*}{Cloud-Edges-Ends} &\multirow{2}{*}{Layers}&\multirow{2}{*}{Latency}&\multirow{2}{*}{Throughput} & \multirow{2}{*}{DP}\\ \\\hline
\multirow{2}{*}{FEPD \citep{ren2021fine}}&\multirow{2}{*}{Cloud-Edges-Ends} &\multirow{2}{*}{Layers}&\multirow{2}{*}{Bandwidth}&Latency & RL\\
&&&&Energy&Early Exiting\\\hline
\multirow{2}{*}{EEOS \citep{chen2021energy}}     &\multirow{2}{*}{Cloud-Edges-Ends}&\multirow{2}{*}{Layers}&\multirow{2}{*}{Latency} &\multirow{2}{*}{Energy}    
& ILP\\
&&&&&PSO-GA\\\hline
\end{tabularx}
\end{table*}

\section{Comparisons}
\label{sec:5}
This section presents the analysis of the DNN partition strategies in terms of two aspects: to refine the metrics of these DNN partition approaches and compare the DNN partition approaches based on these metrics. The characteristics of each algorithm are outlined in detail. 

\subsection{Metrics}

Each DNN partition method is proposed to solve practical problems or improve the reliability of DNN-driven applications. We list some metrics for evaluating the DNN partition algorithms to compare the proposed algorithms.
Qualitative and quantitative indicators are used to measure the DNN partition methods.
\begin{itemize}
\item \emph {\bf The number of optimization performances} includes the number that the DNN partition strategy considered. The optimization performance differs by the DNN partition approaches, such as accuracy, latency, energy, processor usage, and privacy. 
\item \emph {\bf Space complexity} denotes the amount of storage space temporarily occupied by an algorithm when running. It is denoted by $O(\cdot)$. In this study, the space complexity of an algorithm only considers the size of the storage space allocated for local variables during operation. 
\item \emph {\bf Time complexity} denotes the algorithm's running time denoted by $o(\cdot)$. Generally, time complexity relates to the input data and the DNN layers. The weaknesses and strengths are mainly measured in terms of space and time complexity.
\item \emph {\bf Self-adaptability} denotes an ability to execute the DNN partition approach in runtime when the hardware resources and input data vary. 
\item \emph {\bf Generalizability} denotes an indicator for evaluating the overall application value of the DNN partition algorithm. For example, a particular DNN model may be partitionable, but a generalization of this algorithm determines whether the algorithm applies to other DNNs.  In addition, generalization focuses on the universality of the DNN to be split and deployed.
\item \emph {\bf Scalability} indicates whether the algorithm is adaptable when the scenario changes, such as an increase or decrease in the number of sensors and scaling of the number of edge computing nodes. For generalization, scalability focuses on adaptability to the scene.
\end{itemize}
\subsection{ Comparisons }
This subsection presents a comparison of several typical DNN partition approaches. The details are listed in Table~\ref{tab:evaluate}.  
Most DNN partitions have strong generalization, except Pripro~\citep{gao2021pripro}, AppealNet~\citep{li2021appealnet},  ADDA~\citep{wang2019adda}, and Edgent~\citep{li2018edge}. In most in-depth research,  is considered because most DNNs split in given deployment locations. MDPO~\citep{tian2021mobility} and JODS~\citep{duan2021joint} both reduce the delay time of inference. In most DNN partition approaches, there are multiple optimization performances, such as Neurosurgeon~\citep{kang2017neurosurgeon}, Scission~\citep{lockhart2020scission}, and MAHP~\citep{ashouri2020analyzing}.
In particular, Scission~\citep{lockhart2020scission} considers five types of performances; however, it ignores the complexity, generalizability, and other capabilities of the partition model. MAHP~\citep{ashouri2020analyzing} considers more performances and it has greater generalizability compared with Scission. These approaches require many experiments in advance; therefore, it is essential to design lightweight DNN partition approaches. Only a few DNN partition approaches are considered for the time and space complexity, which needs to be further considered. For example, PADCS~\citep{hu2021enable} considers time and space complexity while also having generalizability, scalability, and self-adaptability.

Table~\ref{tab:evaluate} can be extended with as many schemes as one wishes to consider or discuss and is available at  \href{https://github.com/xudi2021/Table-4/blob/main/Table\%204.pdf}{https://github.com/xudi2021/Table-4/blob/main/Table \%204.pdf}.

\begin{table*}[ht]
\footnotesize
\setlength{\tabcolsep}{3pt}
\centering
\caption{\label{tab:evaluate} Comparison of existing works on DNN partition approaches.}

\begin{tabular}{l|cccccc}\hline
\multicolumn{1}{c|}{\multirow{3}{*}{\bf Reference}} & \multicolumn{6}{c}{\bf Metrics}\\
&{\# Optimization} &{Space}&{Time}& \multirow{2}{*}{Generalizability}&\multirow{2}{*}{Scalability}&\multirow{2}{*}{Self-adaptability}\\
&{Performances}&{Complexity}&{Complexity}&&&\\\hline
{DDNNC \citep{zhou2019distributing}}&{1}&{N}&{N}&{N}&{N}&{N}\\\hline
{EdgeLD \citep{xue2020edgeld}}&{1}&{N}&{N}&{N}&{Y}&{Y}\\\hline
{MDPO \citep{tian2021mobility}}&{1}&{N}&{N}&{Y}&{N}&{Y}\\\hline
{EPDNN \citep{shin2019enhanced}}&{1}&{N}&{N}&{Y}&{N}&{Y}\\\hline
{CoopAI \citep{yang2021cooperative}}&{1}&{N}&{Y}&{Y}&{Y}&{N}\\\hline
{JODS \citep{duan2021joint}}&{1}&{N}&{Y}&{Y}&{Y}&{Y}\\\hline
{ADDA \citep{wang2019adda}}&{2}&{N}&{N}&{N}&{N}&{Y}\\\hline
{Edgent \citep{li2018edge}}&{2}&{N}&{N}&{N}&{N}&{Y}\\\hline
{AppealNet \citep{li2021appealnet}}&{2}&{N}&{N}&{N}&{N}&{Y}\\\hline
{Pripro \citep{gao2021pripro}}&{2}&{N}&{N}&{N}&{Y}&{N}\\\hline
{JointDNN \citep{eshratifar2019jointdnn}}&{2}&{N}&{N}&{Y}&{N}&{N}\\\hline
{MeDNN \citep{mao2017mednn}}&{2}&{N}&{N}&{Y}&{N}&{Y}\\\hline
{DADS \citep{hu2019dynamic}}&{2}&{N}&{N}&{Y}&{Y}&{Y}\\\hline
{PANDA \citep{shi2019privacy}}&{2}&{N}&{N}&{Y}&{Y}&{Y}\\\hline
{FEPD \citep{ren2021fine}}&{2}&{N}&{N}&{Y}&{Y}&{Y}\\\hline
{PerDNN \citep{jeong2020perdnn}}&{2}&{N}&{N}&{Y}&{Y}&{Y}\\\hline
{SPSO-GA \citep{chen2021energy}}&{2}&{N}&{Y}&{Y}&{N}&{Y}\\\hline
{QDMP \citep{zhang2020towards}}&{2}&{N}&{Y}&{Y}&{N}&{Y}\\\hline
{TREND-WANT \citep{lin2019distributed}}&{2}&{N}&{Y}&{Y}&{Y}&{N}\\\hline
{JMDP \citep{tang2020joint}}&{2}&{N}&{Y}&{Y}&{Y}&{Y}\\\hline
{JALAD \citep{li2018jalad}}&{2}&{Y}&{Y}&{Y}&{N}&{Y}\\\hline
{PADCS \citep{hu2021enable}}&{2}&{Y}&{Y}&{Y}&{Y}&{Y}\\\hline
{Neurosurgeon \citep{kang2017neurosurgeon}}&{3}&{N}&{N}&{Y}&{N}&{Y}\\\hline
{IONN \citep{jeong2018ionn}}&{3}&{N}&{N}&{Y}&{N}&{Y}\\\hline
{EAIOD \citep{xu2020energy}}&{3}&{N}&{N}&{Y}&{Y}&{Y}\\\hline
{JPDRA \citep{he2020joint}}&{3}&{N}&{Y}&{Y}&{Y}&{Y}\\\hline
{DeepThings \citep{zhao2018deepthings}}&{4}&{N}&{N}&{N}&{N}&{Y}\\\hline
{MCDNN \citep{han2016mcdnn}}&{4}&{N}&{N}&{Y}&{Y}&{Y}\\\hline
{Scission \citep{lockhart2020scission}}&{5}&{N}&{N}&{N}&{N}&{N}\\\hline
{MAHP \citep{ashouri2020analyzing}}&{7}&{N}&{N}&{Y}&{N}&{N}\\\hline

\end{tabular}

\end{table*}

\section{Challenges and Opportunities}
\label{sec:6}
\begin{figure}[ht]
\centering
\includegraphics[width=1.0\textwidth]{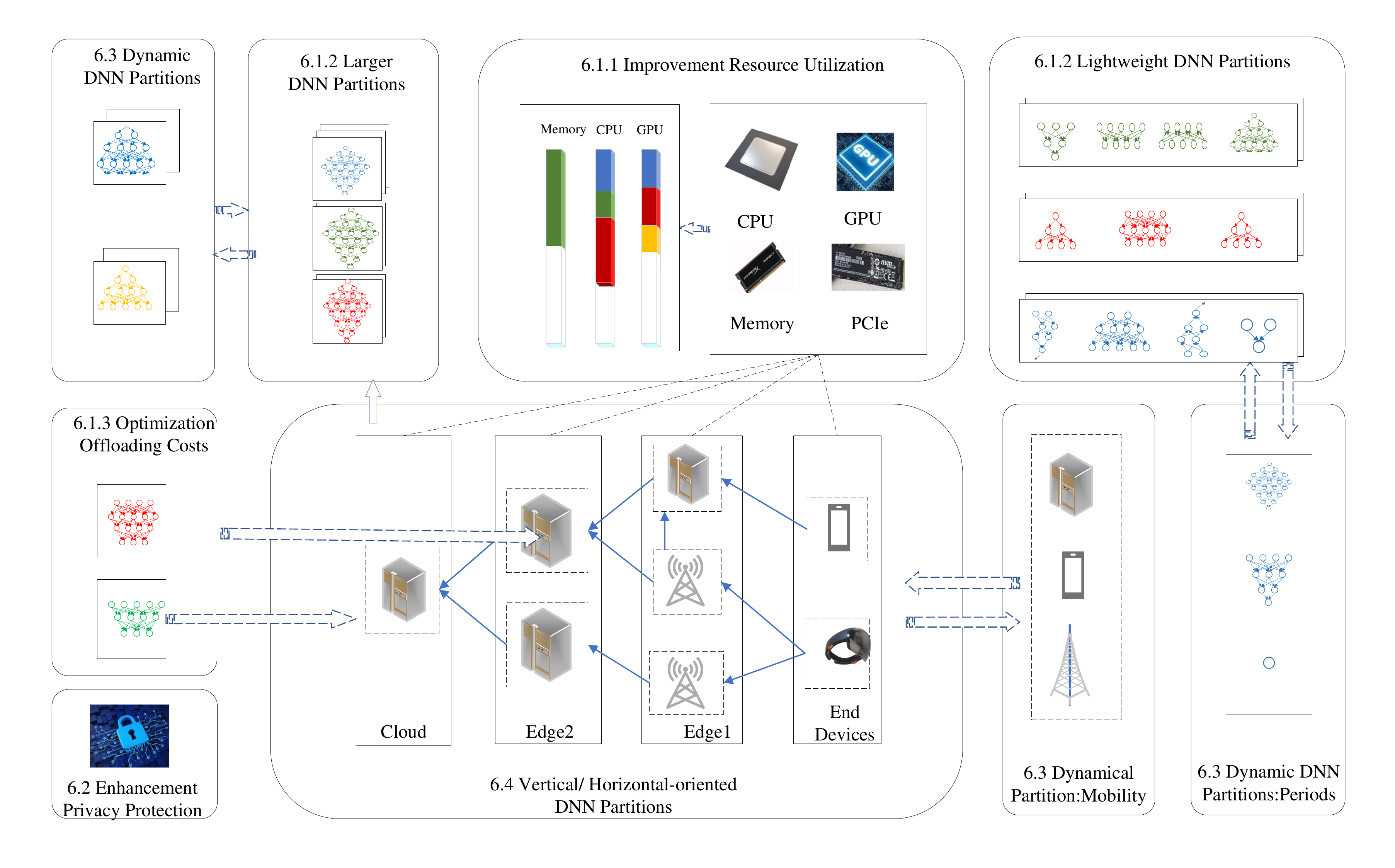}
\caption{\label{fig:challenges} Challenges and opportunities}
\end{figure}

Based on the survey of relevant papers on DNN collaborative inference, we can discuss the limitations of this work, which may provide potential future research directions.
The overall challenges are depicted in detail in Fig.~\ref{fig:challenges}.

\subsection{Consideration of additional factor}

In the aforementioned studies, the DNN partition algorithms considered many factors, such as delay, accuracy, and throughput. However, when establishing the DNN partition optimization model, we considered  other factors, such as resource utilization, DNN magnitude (number of layers), and offloading cost.

\subsubsection{Improved resource utilization}

In most studies, only latency, energy consumption, and accuracy are considered. The delay and energy consumption of the DNN inference are related to the hardware computing power and resources. When the hardware resources are determined in advance, the delay and energy consumption can be obtained. However, the utilization of resources is different from energy consumption and inference delay. The inference time is influenced by sharing a GPU or system resources such as streaming multiprocessors. Therefore, determining the delay and energy consumption by considering only hardware is inaccurate.
However, executing multiple DNN tasks with limited resources results in low memory or processor usage and a congested network. Thus, resource utilization can impact the DNN partition strategy. Therefore, resource utilization and reasonable allocation should be regarded as optimization objectives.

\subsubsection{Additional DNN Partitions}

As the DNN model's scale increases, the partition process generally takes longer. Thus, the DNN partition algorithm must be lightweight and suitable for a more complex DNN model. Recent research has rarely studied the influence of the DNN layers on DNN partition algorithms. As a result, the time and space complexity of the partition algorithms are often ignored. This is a potential future direction for an optimal algorithm: consider the computation complexity.

\subsubsection{Offloading Costs}

Most studies only consider the computing and transmission times of each DNN layer in the total DNN inference latency, assuming simultaneous installation of the entire DNN model on the deployment devices or servers and uploading of the entire layers. This is inappropriate for the emerging edges where the end devices send the intermediate data to the generic servers located at the network's edge. Because the changing offloading edge devices are frequently moved considering the mobile end devices, it is essential to study offloading cost of DNN inference in the dynamic runtime environment.
\subsection{Privacy Protection Concerns}

Although the DNN partition algorithm over the cloud, edge, and end devices, boosts the development of deep learning applications, privacy protection is a significant concern. Sending the DNN intermediate data from edge devices to the cloud is at risk of interception during various stages. The cooperative inference helps to enhance data privacy in DNN-driven applications that employ deep learning models to perform task inference. 
Privacy protection regarding synergy is still in its preliminary stage and requires more research. Therefore, further work could establish a dual goal that considers privacy and accuracy in the constraint of other performance indicators.

\subsection{Dynamic DNN Partitions}

Current IoT applications involve various scenarios. The optimal partition strategy combined with actual scenarios is dynamic; therefore, we need to recalculate the optimal partitions based on the current status of each device being careful to select an interval between recalculations that avoids DNN performance degradation and high overhead.
In addition, end devices and edges have mobile properties; therefore, the deployment location moves, and the number of deployment devices changes. Thus, dynamic deployment locations  must be considered.

\subsection{Vertical- and Horizontal-oriented DNN Partitions }

The E2E-based collaborative computing mode is an essential and promising one, attributable to the support of E2E communication technology. We conclude the DNN partition technology on E2E, called ``horizontal-oriented'' partition. At present, many forms of entertainment, including the famous metaverse (e.g., multiplayer games and AR), are typical multi-end collaboration application scenarios.
In contrast, the computing platform in the ``vertical'' scenario mainly consists of the end device, edges, and cloud. Although the edge nodes to edge nodes oriented DNN partition algorithm has been studied, only a slight gap between the edge node resource and network is assumed. More generally, multi-level edges collaboration partitioning, called ``vertical-oriented'' DNN partition technology is proposed. 
Furthermore, oriented DNN partitions are an excellent attempt at designing DNN partition strategies with regard to the hardware resources and locations.

\section{Conclusion}
\label{sec:7}
This paper provides a comprehensive overview of DNN partition approaches over cloud, edge, and end devices. First, the definition of DNN partition and DNN-based intelligent applications are introduced. Then, the five-dimensional classification framework of DNN partition is described and typical partition approaches are reviewed. Finally, the challenges are listed and several directions for future work are outlined. In summary, DNN partition is a fast-growing research area with numerous challenges and opportunities. We hope that this survey is helpful for understanding state-of-the-art DNN partition research and conducting further research.

\bibliographystyle{unsrtnat}
\bibliography{references} 

\end{document}